\documentclass{ieeetj}
\usepackage{cite}
\usepackage{amsmath,amssymb,amsfonts}
\usepackage{algorithmic}
\usepackage{glossaries}
\usepackage{efbox,graphicx,color}
\usepackage{textcomp}
\usepackage{xcolor}
\usepackage{booktabs}
\usepackage{longtable}
\usepackage[breaklinks]{hyperref} 
\usepackage{soul}
\usepackage{comment}
\hypersetup{hidelinks}
\usepackage{algorithm,algorithmic}

\def\BibTeX{{\rm B\kern-.05em{\sc i\kern-.025em b}\kern-.08em
    T\kern-.1667em\lower.7ex\hbox{E}\kern-.125emX}}
\AtBeginDocument{\definecolor{tmlcncolor}{cmyk}{0.93,0.59,0.15,0.02}\definecolor{NavyBlue}{RGB}{0,86,125}}

\newcolumntype{L}[1]{>{\raggedright\let\newline\\\arraybackslash\hspace{0pt}}m{#1}}
\newcolumntype{C}[1]{>{\centering\let\newline\\\arraybackslash\hspace{0pt}}m{#1}}
\newcolumntype{R}[1]{>{\raggedleft\let\newline\\\arraybackslash\hspace{0pt}}m{#1}}

\def\OJlogo{\vspace{-4pt}$<$Society logo(s) and publication title will appear here.$>$}
\def\seclogo{\vspace{10pt}$<$Society logo(s) and publication title will appear here.$>$}

\def\authorrefmark#1{\ensuremath{^{\textbf{#1}}}}

\newacronym{als}{ALS}{Amyotrophic Lateral Sclerosis}
\newacronym{sand}{SAND}{Speech Analysis for Neurodegenerative Diseases} 
\newacronym{ai}{AI}{Artificial Intelligence}
\newacronym{mfcc}{MFCC}{Mel-Frequency Cepstral Coefficients} 
\newacronym{vocals}{VOC-ALS}{VOiCe signals acquired in Amyotrophic Lateral Sclerosis patients and healthy controls}
\newacronym{hc}{HC}{Healthy Controls} 

\begin{document}
\receiveddate{XX Month, XXXX}
\reviseddate{XX Month, XXXX}
\accepteddate{XX Month, XXXX}
\publisheddate{XX Month, XXXX}
\currentdate{XX Month, XXXX}
\doiinfo{XXXX.2022.1234567}

\markboth{}{Sannino {et al.}}

\title{SAND: The Challenge on Speech Analysis for Neurodegenerative Disease Assessment}

\author{Giovanna Sannino\authorrefmark{1}, Ivanoe De Falco\authorrefmark{1}, Nadia Brancati\authorrefmark{1}, Laura Verde\authorrefmark{2}, Maria Frucci\authorrefmark{1}, Daniel Riccio\authorrefmark{3}, Vincenzo Bevilacqua\authorrefmark{1}, Antonio Di Marino\authorrefmark{1}, Lucia Aruta\authorrefmark{4}, Valentina Virginia Iuzzolino\authorrefmark{5},
  Gianmaria Senerchia\authorrefmark{4}, Myriam Spisto\authorrefmark{6}, and Raffaele Dubbioso\authorrefmark{4}}
  
\affil{National Research Council of Italy (CNR), Institute for High-Performance Computing and Networking (ICAR), Naples, 80131, Italy}
\affil{Department of Mathematics and Physics, University of Campania “Luigi Vanvitelli”, Caserta, 81100, Italy}
\affil{Department of Electrical Engineering and Information Technologies, University of Naples “Federico II”, Naples, 80125, Italy}
\affil{Department of Neurosciences, Reproductive Sciences and Odontostomatology, University of Naples “Federico II”, Naples, 80131, Italy}
\affil{Department of Advanced Biomedical Sciences, University of Naples “Federico II”, Naples, 80131, Italy}
\affil{Department of Psychology, University of Campania “Luigi Vanvitelli”, Caserta, 81100, Italy}
\corresp{Corresponding author: Giovanna Sannino (email: giovanna.sannino@icar.cnr.it).}

\begin{abstract} 
Recent advances in Artificial Intelligence (AI) and the exploration of noninvasive, objective biomarkers, such as speech signals, have encouraged the development of algorithms to support the early diagnosis of neurodegenerative diseases, including Amyotrophic Lateral Sclerosis (ALS). Voice changes in subjects suffering from ALS typically manifest as progressive dysarthria, which is a prominent neurodegenerative symptom because it affects patients as the disease progresses. 
Since voice signals are complex data, the development and use of advanced AI techniques are fundamental to extracting distinctive patterns from them. 
Validating AI algorithms for ALS diagnosis and monitoring using voice signals is challenging, particularly due to the lack of annotated reference datasets. In this work, we present the outcome of a collaboration between a multidisciplinary team of clinicians and Machine Learning experts to create both a clinically annotated validation dataset and the “Speech Analysis for Neurodegenerative Diseases” (SAND) challenge based on it. 
Specifically, by analyzing voice disorders, the SAND challenge provides an opportunity to develop, test, and evaluate AI models for the automatic early identification and prediction of ALS disease progression.
\end{abstract}



\maketitle

\section{INTRODUCTION}
\label{sec:intro}

The \emph{\gls*{sand}} initiative provides an opportunity for the development, testing, and evaluation of \gls*{ai} models for the automatic early diagnosis and longitudinal monitoring of patients with \gls*{als} using voice signals.

Voice changes in subjects suffering from \gls*{als} typically manifest as progressive dysarthria, affecting articulation, prosody, and phonatory control. Several studies have shown that acoustic features can serve as sensitive markers of disease onset and progression, making voice a reliable non-invasive modality for diagnosis and longitudinal monitoring \cite{jothieswari2026dysarthria, dubbioso2024precision}. The availability of robust computational methods for extracting and interpreting acoustical features has opened opportunities for early detection and remote assessment, supporting traditional clinical evaluations.

However, automatically detecting and monitoring neurodegenerative conditions from voice is inherently difficult due to the disorder's heterogeneous manifestations, variable recording conditions, and heterogeneity of affected subjects.

Although voice-based assessment has shown potential, progress is often hindered by limited access to large annotated datasets, by the difficulty of assigning consistent dysarthria severity labels, and by the lack of standardized blind evaluation protocols that mirror realistic deployment conditions.
Moreover, privacy constraints can prevent the release of highly identifying speech material, limiting what can be shared for open challenges.

These factors have led to the development of the SAND challenge \cite{sand_icassp}, and an associated dataset, which aims to stimulate methodological advances in Machine and Deep Learning models (ML/DL) while ensuring fair and reproducible comparisons between different modelling approaches by providing a controlled evaluation framework.

Indeed, the \gls*{sand} challenge includes:
(i) an improved release extending our dataset entitled "\gls*{vocals}"  \cite{vocals_2023,dubbioso2024voice} with more subjects and new data;
(ii) a longitudinal subset enabling disease progression modeling;
(iii) a fully blind, neurologist-annotated test set (released to participants without labels);
(iv) the operational workflow and multidisciplinary practices adopted to ensure data quality, label consistency, and participant anonymity;
and (v) standardized evaluation with an imbalance-robust metric and reference baselines.

The \gls*{sand} challenge attracted significant global interest, reflecting the topic's high relevance, and drew over 800 participants from across five continents, with a predominant representation from academic institutions.

\vspace{-2mm}
\section{DATASET OF THE SAND CHALLENGE}
\label{sec:dataset}

\subsection{Cohort, protocols \& critical aspects}
\label{subsec:cohort_protocols}

The dataset provided by the \gls*{sand} challenge is an improved version of the \gls*{vocals} dataset~\cite{dubbioso2024voice}, with more subjects and with longitudinal recordings. 

It was built through the collaboration of the Department of Neurosciences, Reproductive Sciences and Odontostomatology of the University of Naples ``Federico II'', the Institute for High Performance Computing and Networking (ICAR) of the National Research Council of Italy (CNR), and the University of Campania ``Luigi Vanvitelli''.
This collaboration began in 2021 with the aim of creating a voice signal dataset comprising vowel phonations and syllable repetitions recorded from healthy controls and \gls*{als} patients with varying dysarthria severity, providing annotations by expert neurologists.

A preliminary phase preceded acquisition and annotation: the protocol was reviewed and approved by the medical research Ethics Committee of the ``Federico II'' University Hospital of Naples (Italy), with trial numbers Protocol ID: 100/17/ES01 and 93/2023.
Subjects were enrolled at the ALS Center of the ``Federico II'' University Hospital in Naples between 2022 and 2025 (data collection campaign: January 1, 2022 to June 15, 2025), and included both healthy controls (e.g., caregivers) and patients diagnosed with \gls*{als}.
This phase also included recruitment and informed consent, as well as the collection of demographic and medical history data and clinical assessments; \cite{dubbioso2024voice} reports more details.

Aware of both the effort required to create a large annotated dataset and the complexity of the annotation task due to the difficulty of assessing dysarthria severity, meetings were planned that involved all the figures participating in the study to facilitate knowledge exchange across the different fields.

Before the acquisition process, the ICT team conducted training seminars for the clinical team on the importance of both the quality and the large number of voice signals required to train automatic classification algorithms. 
In particular, it was emphasized that collecting a dataset of heterogeneous signals across the diagnostic spectrum comparable to real-world diagnosis performed by the neurologists, signals with low quality, but with sufficient diagnostic information, had to be kept, while signals that were very noisy had to be removed.
Additionally, in order to ensure a clear understanding of the significant aspects of the problem by the multidisciplinary project teams, the neurologists were instructed by the ICT team on basic principles of the ML/DL to make them understand that the accuracy of an automatic system significantly drops down either if the dataset does not include information-rich examples or it is strongly unbalanced or includes incorrect annotations, or the quality of the signals is very low.

Preliminary analyses and briefings among the different teams required significant time and effort, but enabled proper planning for all subsequent phases of data collection.

\vspace{-4mm}
\subsection{Acquisition setup \& speech recording}
\label{subsec:tasks_acquisition}
\begin{figure}[!ht]
    \centering
    \includegraphics[width=0.65\columnwidth]{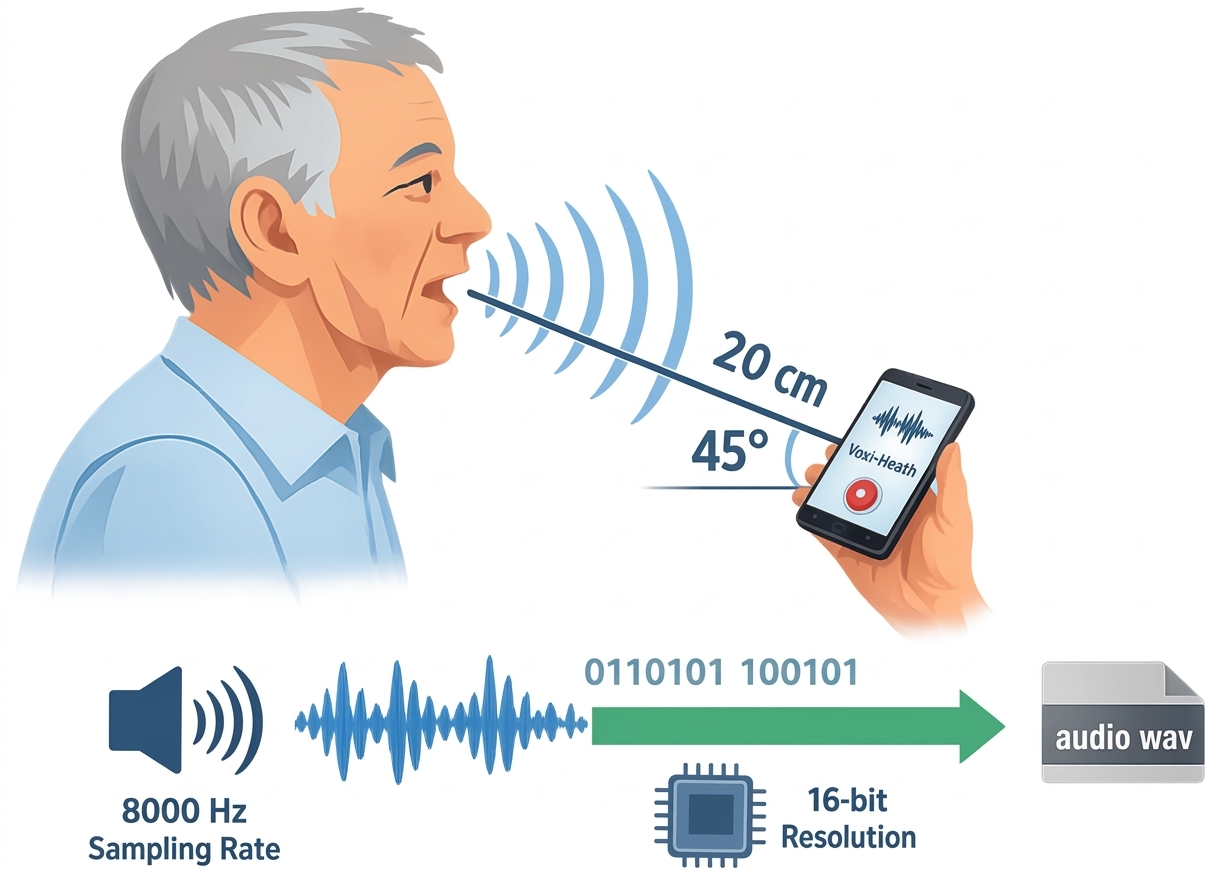}
    \caption{Stylized illustration of the Vox4Health acquisition setup. 
    The smartphone is placed at approximately 20\,cm from the subject’s mouth with an angle of about 45$^\circ$. 
    Audio is recorded via the Vox4Health m-health application and stored as 8\,kHz, 16-bit WAV files.}
    \label{fig:acq_setup}
\vspace{-2mm}
\end{figure}

Audio recordings were collected using the Vox4Health m-health application~\cite{verde2016vox4health,cesari2018voice}. 
During acquisition, the smartphone was positioned at approximately 20\,cm from the subject’s mouth with an angle of about 45$^\circ$, as shown in Fig.~\ref{fig:acq_setup}. 
Signals were digitized at 8\,kHz with 16-bit resolution and stored in \texttt{.wav} format.

To acquire data, subjects were asked to perform different speech exercises and fill out a self-report dysarthria questionnaire. 
The recordings collected for every subject were:
\begin{itemize}\vspace{-1mm}
    \item \textbf{Vowels Vocalizations:} five separate audio files generated by recording each vowel (/a/, /e/, /i/, /o/, and /u/) for at least 5 seconds, maintaining steady volume throughout;
    \item \textbf{Syllables Repetition:} three audio files created as subjects repeated the syllables /pa/, /ta/, and /ka/ as rapidly as possible within a single breath.
\end{itemize}
\vspace{-1mm}
Each patient performed three further tasks: reading text, making a monologue, and repeating the days of the week.
For privacy reasons, we could not include these audio files for the challenge. Indeed, in these recordings, subjects are at high risk of being identified due to the clarity of their voices and speech patterns. Specifically, the monologue, in which subjects describe an image for at least 60 seconds, carries an elevated risk of disclosing sensitive personal information. Consequently, to ensure subject anonymity, these recordings remain confidential and have not been disclosed.

Regarding the demographic data, only age and gender are stored. For patients who performed follow-up, the number of months between the first and last assessments is also stored. Pseudonymization is achieved by replacing subjects' names with unique IDs, and only the medical staff has the link between the pseudonymized and raw data, which are stored locally at the \gls*{als} center. 

Questionnaire, speech, and anonymity details are in \cite{dubbioso2024voice}, while data records will be detailed in the following sections. 

\vspace{-2mm}

\vspace{-4mm}
\subsection{Labeling and Data Quality Assessment}
\label{subsec:labels_consistency}
The \gls*{als} Functional Rating Scale–Revised (ALSFRS-R)~\cite{cedarbaum1999alsfrs} is a 12-item clinical instrument designed to quantify functional impairment in patients with \gls*{als}; in this work, we specifically consider only the speech-related item.

Based on all completed speech tasks and 
questionnaires, an experienced neurologist assigns an ALSFRS-R score to each patient with \gls*{als}. 
This score serves as the dataset label and ranges from 0 (Speech loss) to 4 (No Dysarthria) for \gls*{als} patients; instead, for healthy subjects, it is set to 5.
It is worth noting that no subjects with an ALSFRS-R score of 0 are included in the current dataset; therefore, the lowest ALSFRS-R score is 1 (Severe Dysarthria).

A subsequent verification check of the annotation was required after a discrepancy in the progression of the disease score for follow-up data was detected, and the annotation was collectively revised by two expert neurologists.

\vspace{-2mm}
\subsection{Data format, metadata, splits}
\label{subsec:format_metadata_splits}

\paragraph{Data records.}
The dataset comprises voice signals collected from a cohort of 339 Italian speakers aged between 18 and 90 years.
The cohort consists of:
\begin{itemize}\vspace{-1mm}
    \item \textbf{205 \gls*{als} patients} (121 males, 84 females) presenting varying degrees of dysarthria severity;
    \item \textbf{134 \gls*{hc}} (72 males, 62 females).
\end{itemize}\vspace{-1mm}
Each subject performed $8$ speech tasks ($5$ vowels vocalizations and $3$ syllables repetitions) for a total of $2,712$ voice signals. 
For each subject, we provide basic demographic metadata together with labels derived from the ALSFRS-R scale speech item. 
Subjects are divided into two groups depending on whether a follow-up assessment is available.

For subjects without follow-up (i.e., assessed only once), the metadata include: ID (subject identifier, matching the audio filename prefix), Age, Sex, and the ALSFRS-R Class label, representing dysarthria severity on a scale from 1 to 4 for patients with ALS and 5 for healthy controls.

For subjects with follow-up (i.e., assessed at least twice), the same demographic information is provided, together with additional longitudinal metadata that characterize disease evolution: Months, the time interval between the first and last assessment; ALSFRS-R start, the severity score at the initial assessment; and ALSFRS-R end, the severity score at the final assessment. 

The two groups are not necessarily disjoint, as some subjects may be observed both as single-assessment records and as part of the follow-up cohort when longitudinal recordings are available.

\vspace{-2mm}
\section{SAND CHALLENGE}
\label{sec:challenge}

The SAND Challenge is structured around two complementary tasks, designed to leverage both cross-sectional and longitudinal speech assessments:
\begin{enumerate}\vspace{-1mm}
    \item \textbf{Task 1:} multi-class classification into five classes to estimate the diagnosis and subject’s dysarthria severity;
    \item \textbf{Task 2:} progression prediction across the four ALS severity classes using follow-up data.
\end{enumerate}
\vspace{-1mm}
The dataset for each task is split into training and test sets at the subject level, with an 80/20 split. This ensures that all samples from a single subject are assigned exclusively to either the training or the testing cohort, with no overlap between the two.
Additionally, the split is stratified to preserve a balanced distribution of age, sex, and dysarthria severity (ALSFRS-R score).
All VOC-ALS recordings are included in the training set, together with newly collected data, whereas the test set is entirely new. The involved neurologists annotated all new data.
Test labels are not released to challenge participants because the evaluation is conducted by the organizers under a blind protocol.

As Task~2 relies on longitudinal information, it naturally includes fewer subjects than Task~1, as it is limited to patients who underwent follow-up assessments.

A defining characteristic of both tasks is the pronounced class imbalance within the training set, which mirrors the real-world distribution of clinical data, where severe pathological cases are often harder to recruit than mild or healthy ones.
The class distribution for Task~1 is summarized in Table~\ref{tab:class_task1}. This distribution highlights a significant ``long-tail'' challenge: the Healthy and Non-dysarthric classes (Classes~5 and~4) together constitute over 67\% of the data, while the Severe class accounts for only 2.23\%.

\begin{table}[!ht]
\centering\scriptsize
\caption{Class distribution for Task 1.}
\label{tab:class_task1}
\begin{tabular}{lll}
\toprule
Class & Severity Level & Distribution (\%) \\
\midrule
Class 1 & ALS -- Severe        & 2.23\%  \\
Class 2 & ALS -- Moderate      & 9.55\%  \\
Class 3 & ALS -- Mild          & 20.95\% \\
Class 4 & ALS -- No Dysarthria & 27.94\% \\
Class 5 & Healthy              & 39.33\% \\
\bottomrule
\end{tabular}
\vspace{-2mm}
\end{table}

The class distribution for Task~2, shown in Table~\ref{tab:class_task2}, is more balanced than that of Task~1, though it still exhibits a trend in which asymptomatic or mildly symptomatic states are more prevalent.

\begin{table}[!ht]
\centering\scriptsize
\caption{Class distribution for Task 2.}
\label{tab:class_task2}
\begin{tabular}{lll}
\toprule
Class & Severity Level & Distribution (\%) \\
\midrule
Class 1 & ALS -- Severe        & 13.64\% \\
Class 2 & ALS -- Moderate      & 21.97\% \\
Class 3 & ALS -- Mild          & 28.79\% \\
Class 4 & ALS -- No Dysarthria & 35.60\% \\
\bottomrule
\end{tabular}
\vspace{-2mm}
\end{table}
Since the test set is not released, the official Training Set is further split into a Baseline Training Set (80\%) and a Baseline Validation Set (20\%) to enable method validation and fair comparisons. 

\vspace{-2mm}
\subsection{Data organization}
The dataset is released as a folder tree, with audio files grouped by task and split into training and test. Each split is accompanied by spreadsheet metadata. Within each folder, recordings are stored as \texttt{.wav}. Independent of the task, different spreadsheets are provided depending on the split. 

For the training split, the spreadsheets contain sheets specifying the subjects for the Challenge Training, Baseline Training, and Baseline Validation sets. 
Regarding the information, the spreadsheet provides: (i) \texttt{ID}: subject identifier (matching the audio filename prefix), (ii) \texttt{Age}: age of the subject, (iii) \texttt{Sex}: gender of the subject.
    
Moreover, for Task 1, the spreadsheet includes \texttt{ALSFRS-R} speech score (1--4 for \gls*{als}, 5 for \gls*{hc}), which represents the Class;
Whereas, for Task 2, the spreadsheet also provides the following information: (i) \texttt{Months}: time interval between the first and the last assessment, (ii) \texttt{ALSFRS-R start}: ALSFRS-R speech score at the initial assessment, (iii) \texttt{ALSFRS-R end}, i.e., the ALSFRS-R speech score at the final assessment, which is the Class to predict for Task 2.

Since the test split is a held-out-trials set, the spreadsheet file includes the same information as the training spreadsheets, except for the Class for Task 1 and Task 2.

\vspace{-2mm}
\subsection{Evaluation metric}
\label{subsec:metric_tie}

Performance is assessed using the Averaged F1-score, selected for robustness on unbalanced datasets, and computed on the held-out-trials test set.
For each task, the metric is calculated as:
\begin{equation}
\mathrm{Avg.\ F1Score} =
\frac{1}{|C|}\sum_{c \in C}
\frac{TP_c}{TP_c + \frac{1}{2}(FP_c + FN_c)},
\label{eq:avgf1}
\end{equation}
where $TP_c$ are the true positives, $FP_c$ the false positives, and $FN_c$ the false negatives for label $c$.
Here, $|C|$ is 5 for Task~1 and 4 for Task~2.
Higher values indicate better performance.

\vspace{-2mm}
\subsection{Baselines}
\label{subsec:baselines}

To provide a benchmark for evaluating participant performance, baseline scores were computed using 
state-of-the-art methods 
on the validation split, and the full list of files (hence, subjects) included in both baseline subsets is made available to the challenge participants, even if these scores are not directly comparable to the official blind-test rankings.\newline
\textit{\textbf{Task 1 baseline}}: a Visual Transformer (ViT) has been used as a reference
model \cite{dosovitskiy2020image}. In detail, the 1D waveforms were converted into 2D images by computing spectrograms (Short-Time Fourier Transformations), which were then used as input to the ViT model.
The ViT model leverages self-attention mechanisms to capture global dependencies in acoustic patterns.
The baseline achieved an Avg. F1-score of 0.606 on the validation set. This score reflects the difficulty of a multi-class classification on an unbalanced dataset.\newline
\textit{\textbf{Task 2 baseline}}: we implemented a Partial Decision Tree (PART) algorithm. Unlike traditional decision trees or rule-based learners, PART generates rules by creating partial decision trees at each iteration, which helps in building a more generalized rule set without the need for global optimization.
This baseline model achieved an Avg. F1-Score of 0.583 on the validation dataset. This score reflects the inherent complexity of predicting long-term outcomes from signals, providing a competitive benchmark for participants to improve upon using more sophisticated temporal or deep learning architectures.

\vspace{-2mm}
\subsection{Timeline and participation rules}
\label{subsec:timeline}
The schedule for the SAND challenge was rigorously established in accordance with the rules and deadlines of the
IEEE ICASSP 2026 conference\footnote{\url{https://2026.ieeeicassp.org/}}
within which it was organized.

A dedicated website was created for the SAND challenge\footnote{\url{https://www.sand.icar.cnr.it/}}, where all participants could easily find information on the datasets, the rules for participation, the deadlines, and the organizers' contact details. 
Additionally, through the website, participants could formally register their team for the challenge, access the datasets, submit their algorithms, and calculate the average F1 scores over the test sets.

The initiative officially started on September 1, 2025, marking both the opening of participant registration and the immediate release of the training datasets for both tasks.
On October 1, 2025, the testing datasets for both tasks were released; from that date, participants were allowed a submission window for system results, which concluded on November 20, 2025.
Following internal verification, official results and rankings were announced on December 5, 2025.
High-performing teams were invited to submit 2-page technical papers by January 7, 2026; acceptance notifications were issued on 
January 21, 2026, and the camera-ready deadline was January 28, 2026.
Finally, high-performing teams were invited to present their papers at the IEEE ICASSP conference in Barcelona, Spain, in May 2026.

Regarding the technical setup, participants were given access to the training and test datasets through the dedicated dashboard accessible from the SAND challenge website at predefined times. 
In particular, the training dataset was released one month before the final test set, to give time to define, develop, and validate a model. All model tuning and validation procedures had to be conducted exclusively on the training data. The use or inspection of the test data during model training was strictly forbidden.
The use of external resources, such as pre-trained models or additional datasets, was allowed only if they were publicly available before the start of the challenge and properly cited in the methodology paper. In contrast, the use of any private or proprietary data was strictly prohibited.

The submission protocol allowed teams to compete in either one or both tasks, with a limit of three prediction sets per task, though only the final submission was considered for the official evaluation. To ensure the validity of the results, any incomplete submissions or those containing missing entries were automatically excluded from the ranking process. Finally, each submission had to be accompanied by a two-page methodological description. 

Following the IEEE ICASSP Grand Challenge rules, 
challenge organizers 
invite up to five top-ranked submissions 
to 
present a 2-page 
at IEEE ICASSP 2026. The selection of the top five teams followed a specific distribution logic: the two highest-ranked teams from each task were selected, and the final fifth spot was awarded to the third-ranked team from the task with the highest volume of submissions. 
As detailed in section \ref{sec:challenge}.\ref{subsec:metric_tie}, the ranking was based on the Avg. F1-score. 
In the event of a statistical tie between participants, the organizers reserved the right to use the originality of the proposed methodology as a secondary, decisive criterion.

Although all participants were encouraged to share their code publicly, to verify the top methods and validate the final ranking of the winners, the highest-performing teams for each task were required to provide executable files or notebooks to reproduce their results.

\vspace{-3mm}
\section{PARTICIPANTS}
\label{sec:participants}
\begin{figure}
\centerline{\includegraphics[width=\linewidth]{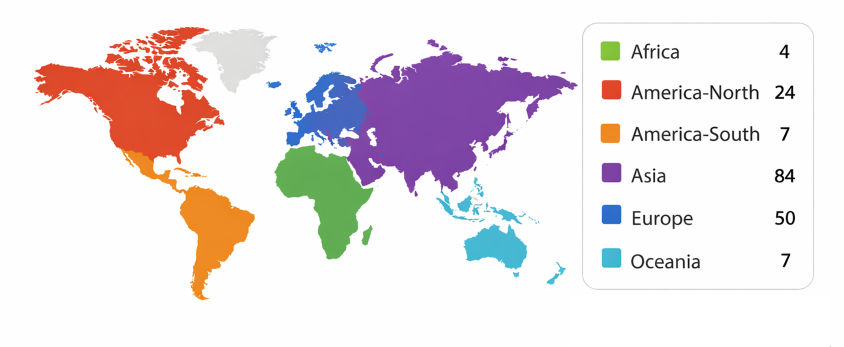}}\vspace{-4mm}
\caption{Number of teams registered for the SAND Challenge for each continent.\label{fig:NumberTeamsContinent}}
\vspace{-4mm}
\end{figure}

The SAND Challenge successfully engaged a diverse global community, attracting 176 teams from all continents, totalling about 850 participants. As shown in Figure \ref{fig:NumberTeamsContinent}, Asia led participation with 84 teams, representing nearly half of the total cohort. Europe followed with 50 teams, while North America contributed 24 teams to the initiative. Participation was further rounded out by Oceania and South America, each with 7 teams, and Africa, with 4 teams.

\begin{figure*}[!ht]
\centerline{\includegraphics[width=0.9\linewidth]{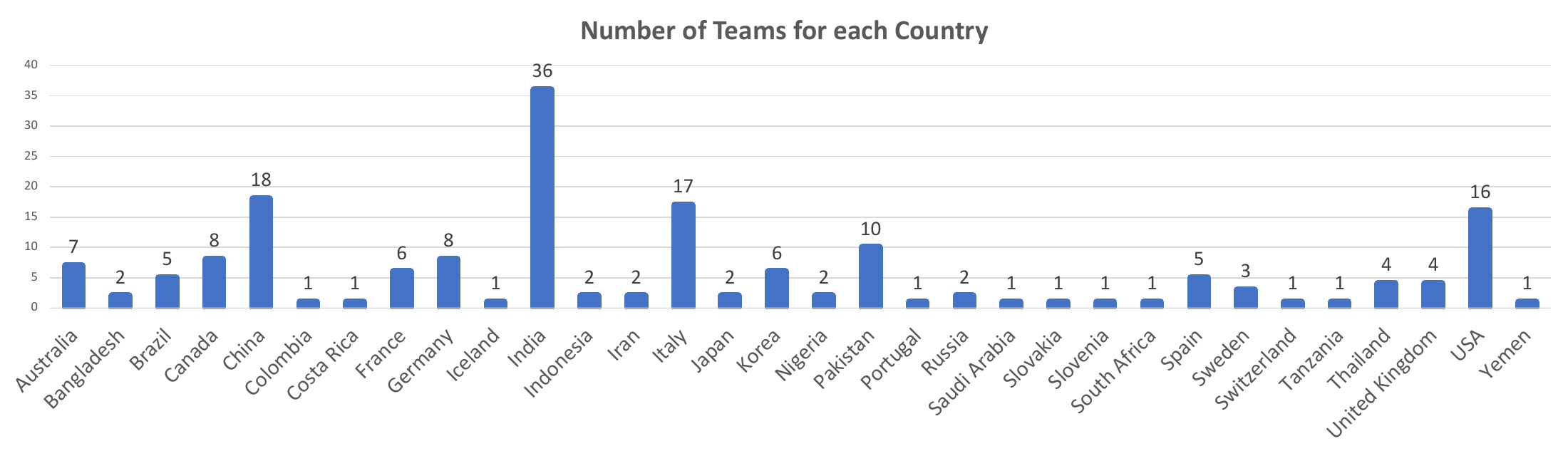}}\vspace{-4mm}
\caption{Number of Teams registered for the SAND Challenge for each country.\label{fig:NumberTeamsCountry}}
\vspace{-2mm}
\end{figure*}

At the national level, India stood out as the most active country with 36 teams, followed by China (18 teams), Italy (17 teams), and the USA (16 teams). Other significant contributions came from Pakistan (10 teams), as well as Canada and Germany, each fielding 8 teams. The complete distribution of registered participating teams across countries is shown in Figure \ref{fig:NumberTeamsCountry}. 

Although some registered teams did not complete the challenge, participation in the individual tasks remained substantial: 231 and 84 participants, for Task 1 and Task 2, respectively. Detailed information on actual participation in each task is provided in the following subsections. 

\vspace{-2mm}
\subsection{Participation in Task 1}
\label{sebsec:statistics-task-1}
A total of 231 participants from 61 teams took part in Task 1. The majority (205) were affiliated with academia, while 15 participants came from industry, and the remaining 11 were engaged in other activities.
Most participants were from Asia (98), followed by Europe (90) and North America (28). In terms of team composition, the number of participants per team ranged from 1 to 14, with an average team size of 3.79 participants. Notably, four teams were composed of participants from multiple countries/continents.
 
\begin{figure*}[!ht]\vspace{-2mm}
\centerline{\includegraphics[width=0.9\linewidth]{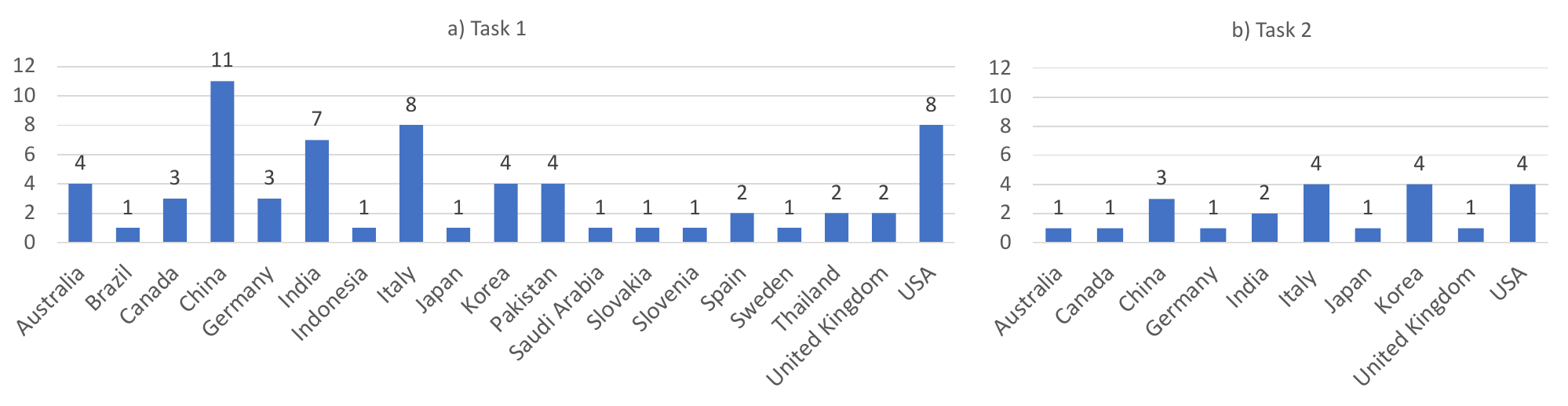}}\vspace{-2mm}
\caption{Number of Teams Participating for each country in a) Task 1, and b) Task 2.\label{fig:Teams_Task1_and2}}
\end{figure*}

In terms of team distribution, most teams were based in China, followed by the United States of America and Italy, as shown in Figure \ref{fig:Teams_Task1_and2}.a. 

Considering the continental distribution of teams, the majority originated from Asia (
$\sim$48\%), followed by Europe (
$\sim$26\%) and the Americas (
$\sim$17\%).

\vspace{-2mm}
\subsection{Participation in Task 2}
\label{sebsec:statistics-task-2}
Regarding Task 2, the number of participants decreased to 84 affiliated with 22 teams, most of whom were from academia (68), while the remainder were from industry (11) and other activities (5). In terms of continents, the largest number of researchers came from Asia (37), followed by Europe (36). At the country level, Italy contributed 28 participants, followed by Korea with 20.
When team sizes are taken into account, the number of participants per team ranged from 1 to 11, with an average team size of 3.818.
This indicates the presence of several one-person teams (six, in fact), as well as one large team with 11 members; nevertheless, the average is around 4 researchers per team.

Figure \ref{fig:Teams_Task1_and2}.b shows the number of teams for each country: the highest participation was from Italy, Korea, and the USA, with four teams each. In total, 11 countries
were represented.

Concerning the number of teams per continent, the highest participation was from Asia, with 11 teams, followed by Europe, with 6. In Figure \ref{fig:Teams_Task1_and2}, the total number of participating teams is 23 rather than 22; this is because one team comprised researchers from both Pakistan and the United Kingdom, thereby contributing to both Asia and Europe.

\vspace{-2mm}
\section{RESULTS}
\label{sec:results}
Sixty-one teams submitted their predictions for Task 1. 
However, only fifty-two of these teams followed the correct submission procedure and were therefore included in the ranking.
The results obtained by each team are ranked in decreasing order of Avg. F1-score in Table \ref{tab:results-task1}. 

As for Task 2, the Avg. F1-score values obtained by the twenty-two teams that submitted their prediction sets are reported in decreasing order in Table \ref{tab:results-task2}. 

Table \ref{tab:statis_results} summarizes the distribution and variability of the results for the two tasks. 
The first row shows the statistics for Task 1: the results ranged from 0.0564 to 0.6079, with a mean of 0.4004, a median of 0.4005, and a standard deviation of 0.1427. 
The second row refers to Task 2; here, the results ranged from 0.3069 to 0.5794, with a mean of 0.4463, a median of 0.4351, and a standard deviation of 0.0825. The comparison of the two standard deviation values suggests that, on average, the results obtained by the different methods on Task 2 were more similar to each other than those obtained on Task 1. 

Following the ICASSP Grand Challenge rules, we invited only the top five ranked teams to 
present a 2-page paper 
at ICASSP 2026, i.e., proclaimed these top five ranked teams as winners.
As 
mentioned in Section \ref{subsec:timeline}, the selection process followed a specific distribution logic: the two highest-ranked teams from each task were automatically selected. The fifth and final spot was 
awarded to the third-ranked team from the task with the highest volume of submissions.
Since Task 1 saw higher participation than Task 2 (61 vs 22), the final selection comprised the top three teams from Task 1 and the top two teams from Task 2, 
based on the achieved F1-score. 

In the following subsections, we provide a detailed description of the methodologies proposed by the winning teams, including a discussion of the specific approaches adopted for each task.

\begin{table}[!ht]\tiny\centering
\renewcommand\arraystretch{1.2}
\caption{Task 1 rankings.}
\begin{tabular}{ccc|ccc}
\hline
\textbf{Rank} & \textbf{Team Name} & \textbf{Avg.\ F1} &
\textbf{Rank} & \textbf{Team Name} & \textbf{Avg.\ F1} \\
\hline
1°  & TUKE               & 0.6079 & 27° & CLT                & 0.3937 \\
2°  & UTL                & 0.6005 & 28° & CAU                & 0.3937 \\
3°  & PRIME              & 0.5945 & 29° & SMTIH              & 0.3917 \\
4°  & RGTRGT             & 0.5849 & 30° & BPGC               & 0.3813 \\
5°  & CCNYNEURO          & 0.5813 & 31° & PHOFI              & 0.3793 \\
6°  & OHTSUKI            & 0.5796 & 32° & AICV               & 0.3757 \\
7°  & UTAUSTIN           & 0.5768 & 33° & UHL                & 0.3725 \\
8°  & SARWANALI          & 0.5613 & 34° & DSPLABMARIBOR      & 0.3645 \\
9°  & PASSIONAI          & 0.5558 & 35° & IITPATNA           & 0.3634 \\
10° & AURA               & 0.5437 & 36° & TEAMTAG            & 0.3629 \\
11° & TKB                & 0.5430 & 37° & CSCU               & 0.3623 \\
12° & SLEEPERS           & 0.5320 & 38° & GTMN               & 0.3596 \\
13° & GTIUNISS           & 0.5301 & 39° & JLEE               & 0.3398 \\
14° & ISDS               & 0.5116 & 40° & GTMUVIGO           & 0.3353 \\
15° & PATHOLOGICALSPEECH & 0.4820 & 41° & TSY                & 0.3078 \\
16° & MOCHA              & 0.4815 & 42° & STAR               & 0.2969 \\
17° & QLN                & 0.4804 & 43° & ECHOPATH           & 0.2870 \\
18° & CAB                & 0.4794 & 44° & IMATI              & 0.2629 \\
19° & SSS                & 0.4790 & 45° & GISPHEU            & 0.2404 \\
20° & MBS                & 0.4751 & 46° & FBK                & 0.2025 \\
21° & UOS                & 0.4413 & 47° & WTB                & 0.1996 \\
22° & SMARTVOICE         & 0.4231 & 48° & VAPMR              & 0.1888 \\
23° & SAGI               & 0.4175 & 49° & TCSSPEECH          & 0.1362 \\
24° & SPAGHETTIINHALERS  & 0.4159 & 50° & MACEWANVOICES      & 0.0992 \\
25° & CASALAB            & 0.4137 & 51° & ALWAYSMAKEIMPACT   & 0.0656 \\
26° & THR                & 0.4073 & 52° & NETSENSE           & 0.0564 \\
\hline
\end{tabular}\vspace{-4mm}
\label{tab:results-task1}
\end{table}

\begin{table}[!ht]\tiny\centering
\renewcommand\arraystretch{1.2}
\caption{Task 2 rankings.}
\begin{tabular}{ccc}
\hline
\textbf{Rank} & \textbf{Team Name } & \textbf{Avg.F1-score} \\
\hline
1°   & ISDS               & 0.5794   \\
2°   & OHTSUKI            & 0.5637   \\
3°   & JLEE               & 0.5612   \\
4°   & CAU                & 0.5437   \\
5°   & SPAGHETTIINHALERS  & 0.5401   \\
6°   & PATHOLOGICALSPEECH & 0.5278   \\
7°   & ECHOPATH           & 0.4994   \\
8°   & CAB                & 0.4870   \\
9°   & CLT                & 0.4791   \\
10°  & AICV               & 0.4552   \\
11°  & CCNYNEURO          & 0.4408   \\
12°  & SMARTVOICE         & 0.4294   \\
13°  & AURA               & 0.4185   \\
14°  & KIE                & 0.4097   \\
15°  & HNDX               & 0.3879   \\
16°  & PASSIONAI          & 0.3815   \\
17°  & SARWANALI          & 0.3795   \\
18°  & ARCOLAB            & 0.3728   \\
19°  & TKB                & 0.3673   \\
20°  & MOCHA              & 0.3667   \\
21°  & MBS                & 0.3207   \\
22°  & TEAMTAG            & 0.3069  \\
\hline
\end{tabular}
\label{tab:results-task2}
\vspace{-2mm}
\end{table}

\begin{table}[!ht]
  \scriptsize\centering
  \renewcommand\arraystretch{1.2}
  \caption{Statistical analysis of the results.}
    \begin{tabular}{L{0.8cm}C{1cm}C{1cm}C{1cm}C{1cm}C{1.2cm}}
    \toprule
          & \textbf{Max} & \textbf{Min} & \textbf{Mean} & \textbf{Median} & \textbf{Std. Dev.} \\
    \midrule
    \textit{\textbf{Task 1}} & 0.6079 & 0.0564 & 0.4004 & 0.4005 & 0.1427 \\
    \textit{\textbf{Task 2}} & 0.5794 & 0.3069 & 0.4463 & 0.4351 & 0.0825 \\
    \bottomrule
    \end{tabular}
    \label{tab:statis_results}
\vspace{-4mm}
\end{table}

\vspace{-3mm}
\subsection{Task 1 Top Teams' Approaches}
In the following, we report the results of the three top teams' approaches for Task 1.

\vspace{-5mm}
\subsubsection{\textbf{TUKE Team}}
The TUKE Team from the Technical University of Košice (Slovakia) achieved first place in Task 1 \cite{tuke_icassp} by proposing a hierarchical, coarse-to-fine framework built on the Whisper-Turbo foundation model developed by OpenAI \cite{radford2023robust}.
Audio recordings are represented as mel-spectrograms, next a three-stage progressive adaptation strategy is applied: (1) broad pre-adaptation on heterogeneous neurodegenerative speech datasets (including Parkinson’s and Alzheimer’s) to learn disorder-agnostic dysarthria representations, using gradual encoder unfreezing; (2) coarse-to-fine fine-tuning on the ALS domain, initially merging the five severity levels into two macro-classes; and (3) specialization into two expert models, one focused on “no dysarthria” discrimination and the other on differentiating dysarthric severity levels, with partial encoder fine-tuning to capture subtle severity cues.

Finally, sample-level predictions were aggregated at the subject level via majority voting, yielding a winning Avg. F1-score of $0.6079$ on the test set.

\vspace{-5mm}
\subsubsection{\textbf{UTL Team}}
The runner-up was
the UTL team from the University of Texas at Austin (USA), which proposed a hierarchical framework 
merging
an Audio Spectrogram Transformer (AST) \cite{gong2021ast} 
and
a Multiple Instance Learning (MIL) strategy \cite{utl_icassp}.

Their pipeline extracts Log-Mel spectrograms from eight preprocessed audio recordings (five sustained vowels and three syllable repetitions per subject) and feeds them into an AST encoder based on a Vision Transformer architecture to learn spectral--temporal representations. An attention-based MIL module then aggregates multiple recordings per subject, revealing that the model assigns higher importance scores to rhythmic syllable repetition tasks than to sustained vowels when distinguishing between moderate and mild dysarthria. 
The obtained Avg. F1-score was $0.6005$ on the test set. 
Moreover,
several interpretability methods were applied (e.g. Saliency, Occlusion, Gradient Shap), 
among which the Occlusion model confirmed that the approach focuses on high-frequency instability patterns 
typical of dysarthric speech.

\vspace{-5mm}
\subsubsection{\textbf{PRIME Lab Team}}
PRIME Lab Team from the Université de Moncton (Canada) took third, proposing an approach based on visual transformation and ensemble learning \cite{prime_icassp}.

Their method represents the eight phonations of each patient within a single input using a spectrogram mosaic, allowing deep vision models to process all recordings simultaneously. Among several audio representations, the Constant-Q Transform (CQT) \cite{sun2024constant} proved to be the most effective for capturing dysarthric markers. The resulting spectrograms were zero-padded to manage variable durations and combined using different mosaic layouts (vertical stacking, horizontal sequence, or a $3\times3$ grid).

The mosaics were then classified using pretrained vision architectures such as DenseNet-121 and ConvNeXt-Base, with additional data augmentation (noise injection, time shifting, and resampling). Finally, an ensemble with majority voting across the best models was implemented, achieving an Avg. F1-score of 0.5945 on the test set.

\vspace{-5mm}
\subsubsection{\textbf{Discussion}}
Despite architectural differences, all methods rely on time–frequency representations of speech signals, confirming the central role of spectrogram-based features in dysarthria analysis. 
Both the TUKE and UTL teams adopt Mel-based representations (Mel-spectrograms and log-Mel spectrograms, respectively). PRIME Lab also performs some experiments using Mel-spectrograms, even if alternative transforms are explored, such as the Constant-Q Transform (CQT), achieving competitive performance. This convergence suggests that transforming raw audio into structured spectral representations is the prevailing approach for capturing clinically relevant speech patterns. 

Regarding the adopted models, the UTL and PRIME Lab teams implement methods to handle multiple recordings per subject and aggregate information. In detail, UTL adopts a MIL-based aggregation scheme, and PRIME Lab integrates multiple inputs via mosaic representations followed by ensemble learning. On the other hand, the TUKE team employs a cross-to-fine pipeline that effectively aggregates information across stages. 
Given the variability across speech recordings, aggregation mechanisms are crucial, whether applied to combine multiple inputs, i.e., UTL and PRIME Lab's approaches, or to aggregate class predictions, as demonstrated by the TUKE approach. 

However, the comparison reveals three complementary strategies of the approaches:
\begin{itemize}\vspace{-1mm}
\item TUKE: Speech foundation model adaptation. The method focuses on transfer learning and domain adaptation, leveraging a large pretrained speech model (Whisper-Turbo) and refining it through a carefully designed coarse-to-fine training strategy. Their use of task decomposition into expert models is particularly performing for handling class imbalance and subtle severity distinctions.
\item UTL: Transformer-based modeling with attention-driven aggregation. The method emphasizes representation learning and interpretability, combining an AST with attention-based MIL. This allows the model not only to aggregate multiple inputs but also to identify which speech tasks are most informative, highlighting the importance of rhythmic syllable repetitions.
\item PRIME Lab: Vision-based transformation with ensemble learning. The use of mosaic representations, as a vision-oriented strategy, and ensemble learning prioritize robustness and leverage strong pretrained vision backbones rather than task-specific architectural design.
\end{itemize}

\vspace{-5mm}
\subsection{Task 2 Top Teams' Approaches}
In the following, we report the results of the two top teams' approaches for Task 2.

\vspace{-5mm}
\subsubsection{\textbf{ISDS Team}}
For Task 2, the best result was achieved by the ISDS team from Sogang University in Seoul (South Korea), which proposed a multi-feature architecture combining self-supervised speech embeddings, acoustic features, and metadata to predict \gls*{als} progression \cite{isds_icassp}.
Their approach processes raw audio using a fine-tuned WavLM encoder, whose embeddings are fused with \gls*{mfcc} and patient metadata, such as age, gender, task type, and initial severity score. Both WavLM and \gls*{mfcc} features undergo a projection-then-pooling strategy, in which frame-level projections are computed before temporal mean pooling to preserve discriminative information.

The resulting features are concatenated and fed into a multi-layer classifier with a residual connection that directly incorporates the baseline severity score. Final patient-level predictions are obtained through majority voting across the outputs of the three best-performing models, improving robustness against overfitting.
The proposed method achieved an Avg. F1-score of 0.5794 on the test set.

\vspace{-5mm}
\subsubsection{\textbf{OHTSUKI Team}}
Second place in Task 2 was achieved by the OHTSUKI team from Keio University in Tokyo (Japan), which proposed a syllable-level acoustic modeling framework based on a stage-aware Transformer architecture \cite{othsuki-icassp}.
Their approach focuses on syllable repetition tasks, in which recordings are segmented into /pa/, /ta/, and /ka/ units using the Sylber tool \cite{cho2024sylber}. From these segments, GeMAPS \cite{eyben2015geneva} acoustic descriptors are extracted with OpenSMILE \cite{eyben2013recent} and filtered through Mutual Information analysis to retain the most discriminative features for each disease stage.

The key idea is a stage-aware learning strategy that routes patients to specialized Transformer classifiers based on their initial ALSFRS-R severity level, enabling the model to learn progression patterns tailored to each stage. Acoustic features are combined with metadata such as gender, normalized age, and time intervals 
within 
a fixed-length representation of 60 syllables, enabling 
capturing fine-grained progression dynamics. 
The Avg. F1-score on the test set was 0.5637.

\vspace{-5mm}
\subsubsection{\textbf{Discussion}}
Both approaches integrate acoustic features with patient metadata, i.e., 
the initial severity score, along with demographic and clinical information (e.g., age, gender) and speech-derived features, highlighting the importance of multimodal data for improving prediction performance. Another shared aspect is the focus on robust feature selection and representation. ISDS uses a projection-then-pooling strategy to preserve discriminative information from embeddings, whereas OHTSUKI applies Mutual Information to retain only the most informative acoustic descriptors.
However, despite these similarities, the two methods differ substantially in their design philosophy:
\begin{itemize}\vspace{-1mm}
\item ISDS: Multimodal feature fusion with ensemble learning. Self-supervised embeddings, handcrafted acoustic features and metadata are merged into a unified representation and processed by a single multi-layer classifier, with an additional residual connection incorporating baseline severity. Finally, an ensemble learning improves the robustness.
\item OHTSUKI: Stage-aware syllable-level Transformer modeling. The approach focuses on syllable-level modeling. Moreover, a stage-aware learning strategy is adopted: audio recordings are routed to different Transformer models depending on their initial severity level, allowing 
learning
stage-specific progression dynamics.
\end{itemize}\vspace{-1mm}

\vspace{-4mm}
\section{CONCLUSIONS}
\label{sec:conclusions}
Computational analysis of voice signals has the potential to enable novel applications for speech technology in the diagnosis and monitoring of neurodegenerative diseases, including Amyotrophic Lateral Sclerosis (ALS). 

The {\it Speech Analysis for Neurodegenerative Diseases (SAND)} Challenge provides an opportunity to investigate alternative approaches to neurological and clinical evaluation for \gls*{als} diagnosis and monitoring. To this aim, the SAND challenge provided a dataset including repeated occurrences of vocalisations/syllables from participants, both patients with different levels of dysarthria 
and healthy controls, with imbalances in sex and age. The challenge was hosted within the ICASSP 2026 Conference and consisted of two tasks, one related to diagnosis and the other to progression. 

Despite the complexity of the tasks, the challenge received a large number of high-quality solutions. Indeed, 176 teams from around the world registered, and some participated in both tasks. For Task 1, the best-performing method achieved an Avg. F1-score of 0.6079 in classifying five classes to estimate the diagnosis and the subject’s dysarthria severity. Regarding Task 2, an Avg. F1-score of 0.5794 was obtained by the best-performing method to predict disease progression across the four ALS severity classes using follow-up data.

The participating teams proposed a wide range of ML/DL methodologies, further underscoring AI's potential to address the challenges posed by \gls*{als}. It is the organizers’ hope that the comprehensive analysis presented herein will motivate further challenges in the diagnosis and monitoring of neurodegenerative diseases by analyzing voice signals, with clear benefits for our society.



\vspace{-2mm}
\section*{ACKNOWLEDGMENT}
The authors acknowledge the D34Health project (PNC0000001 - CUP B83C22006120001), within the Italian PNC-PNRR, financed by the European Union; and the HISENSE project (FIS-2024-06824 - B53C25003830001), funded by the Italian MUR within the FIS3 call.

\bibliographystyle{IEEEtran}

\begin{thebibliography}{10}
\providecommand{\url}[1]{#1}
\csname url@samestyle\endcsname
\providecommand{\newblock}{\relax}
\providecommand{\bibinfo}[2]{#2}
\providecommand{\BIBentrySTDinterwordspacing}{\spaceskip=0pt\relax}
\providecommand{\BIBentryALTinterwordstretchfactor}{4}
\providecommand{\BIBentryALTinterwordspacing}{\spaceskip=\fontdimen2\font plus
\BIBentryALTinterwordstretchfactor\fontdimen3\font minus \fontdimen4\font\relax}
\providecommand{\BIBforeignlanguage}[2]{{%
\expandafter\ifx\csname l@#1\endcsname\relax
\typeout{** WARNING: IEEEtran.bst: No hyphenation pattern has been}%
\typeout{** loaded for the language `#1'. Using the pattern for}%
\typeout{** the default language instead.}%
\else
\language=\csname l@#1\endcsname
\fi
#2}}
\providecommand{\BIBdecl}{\relax}
\BIBdecl

\bibitem{jothieswari2026dysarthria}
J.~Jothieswari and S.~Suguna, ``Dysarthria speech disorder detection: A recent review,'' in \emph{International Conference on Hybrid Intelligence: Theories and Applications}.\hskip 1em plus 0.5em minus 0.4em\relax Springer, 2026, pp. 173--187.

\bibitem{dubbioso2024precision}
R.~Dubbioso, M.~Spisto, L.~Verde, V.~V. Iuzzolino, G.~Senerchia, G.~De~Pietro, I.~De~Falco, and G.~Sannino, ``Precision medicine in {ALS}: Identification of new acoustic markers for dysarthria severity assessment,'' \emph{Biomedical Signal Processing and Control}, vol.~89, p. 105706, 2024.

\bibitem{sand_icassp}
G.~Sannino, I.~De~Falco, N.~Brancati, L.~Verde, M.~Frucci, D.~Riccio, V.~Bevilacqua, A.~Di~Marino, L.~Aruta, V.~V. Iuzzolino, G.~Senerchia, M.~Spisto, and R.~Dubbioso, ``The speech analysis for neurodegenerative diseases challenge,'' in \emph{ICASSP 2026--2026 IEEE International Conference on Acoustics, Speech and Signal Processing ({ICASSP})}.\hskip 1em plus 0.5em minus 0.4em\relax IEEE, 2026.

\bibitem{vocals_2023}
\BIBentryALTinterwordspacing
G.~Sannino, I.~De~Falco, V.~V. Iuzzolino, E.~Salvatore, G.~Senerchia, M.~Spisto, L.~Verde, and R.~Dubbioso, ``{VOC-ALS} database, {VO}i{C}e signals acquired in amyotrophic lateral sclerosis patients,'' 2023. [Online]. Available: \url{https://repo-prod.prod.sagebase.org/repo/v1/doi/locate?id=syn53009474&type=ENTITY}
\BIBentrySTDinterwordspacing

\bibitem{dubbioso2024voice}
R.~Dubbioso, M.~Spisto, L.~Verde, V.~V. Iuzzolino, G.~Senerchia, E.~Salvatore, G.~De~Pietro, I.~De~Falco, and G.~Sannino, ``Voice signals database of {ALS} patients with different dysarthria severity and healthy controls,'' \emph{Scientific Data}, vol.~11, no.~1, p. 800, 2024.

\bibitem{verde2016vox4health}
L.~Verde, G.~De~Pietro, and G.~Sannino, ``{Vox4Health}: Preliminary results of a pilot study for the evaluation of a mobile voice screening application,'' in \emph{International Symposium on Ambient Intelligence}.\hskip 1em plus 0.5em minus 0.4em\relax Springer, 2016, pp. 131--140.

\bibitem{cesari2018voice}
U.~Cesari, G.~De~Pietro, E.~Marciano, C.~Niri, G.~Sannino, and L.~Verde, ``Voice disorder detection via an m-health system: design and results of a clinical study to evaluate vox4health,'' \emph{BioMed research international}, vol. 2018, no.~1, p. 8193694, 2018.

\bibitem{cedarbaum1999alsfrs}
J.~M. Cedarbaum, N.~Stambler, E.~Malta, C.~Fuller, D.~Hilt, B.~Thurmond, A.~Nakanishi, {Bdnf Als Study Group} \emph{et~al.}, ``The {ALSFRS-R}: A revised {ALS} functional rating scale that incorporates assessments of respiratory function,'' \emph{Journal of the Neurological Sciences}, vol. 169, no. 1--2, pp. 13--21, 1999.

\bibitem{dosovitskiy2020image}
A.~Dosovitskiy, L.~Beyer, A.~Kolesnikov, D.~Weissenborn, X.~Zhai, T.~Unterthiner, M.~Dehghani, M.~Minderer, G.~Heigold, S.~Gelly \emph{et~al.}, ``An image is worth 16x16 words: Transformers for image recognition at scale,'' \emph{arXiv preprint arXiv:2010.11929}, 2020.

\bibitem{tuke_icassp}
S.~Hresko, M.~Hires, J.~Stas, and P.~Drotar, ``A hierarchical coarse-to-fine {Whisper} adaptation framework for {ALS} dysarthria severity estimation,'' in \emph{ICASSP 2026--2026 IEEE International Conference on Acoustics, Speech and Signal Processing ({ICASSP})}.\hskip 1em plus 0.5em minus 0.4em\relax IEEE, 2026.

\bibitem{radford2023robust}
A.~Radford, J.~W. Kim, T.~Xu, G.~Brockman, C.~McLeavey, and I.~Sutskever, ``Robust speech recognition via large-scale weak supervision,'' in \emph{Proceedings of the 40th International Conference on Machine Learning ({ICML})}.\hskip 1em plus 0.5em minus 0.4em\relax PMLR, 2023, pp. 28\,492--28\,518.

\bibitem{gong2021ast}
Y.~Gong, Y.-A. Chung, and J.~Glass, ``{AST}: Audio spectrogram transformer,'' \emph{arXiv preprint arXiv:2104.01778}, 2021.

\bibitem{utl_icassp}
P.~A. Alba~Diaz, A.~A. Kedilaya, R.~Kolm, and J.~Robertson, ``Audio spectrogram transformer and multiple instance learning for amyotrophic lateral sclerosis severity classification,'' in \emph{ICASSP 2026--2026 IEEE International Conference on Acoustics, Speech and Signal Processing ({ICASSP})}.\hskip 1em plus 0.5em minus 0.4em\relax IEEE, 2026.

\bibitem{prime_icassp}
M.~A. Blais and M.~A. Akhloufi, ``{ALS} detection from phonation audio using spectrogram mosaics and ensemble deep learning,'' in \emph{ICASSP 2026--2026 IEEE International Conference on Acoustics, Speech and Signal Processing ({ICASSP})}.\hskip 1em plus 0.5em minus 0.4em\relax IEEE, 2026.

\bibitem{sun2024constant}
G.~Sun and L.~Wang, ``Constant-{Q} transform for audio-visual dysarthria severity assessment,'' in \emph{2024 IEEE 14th International Symposium on Chinese Spoken Language Processing ({ISCSLP})}.\hskip 1em plus 0.5em minus 0.4em\relax IEEE, 2024, pp. 146--150.

\bibitem{isds_icassp}
I.~Lee, T.~Jeong, M.~Han, Y.~Lee, and M.~W. Koo, ``{WavLM}-based feature fusion with metadata for {ALS} severity prediction,'' in \emph{ICASSP 2026--2026 IEEE International Conference on Acoustics, Speech and Signal Processing ({ICASSP})}.\hskip 1em plus 0.5em minus 0.4em\relax IEEE, 2026.

\bibitem{othsuki-icassp}
Y.~Tamura, M.~Bouazizi, and T.~Ohtsuki, ``Syllable-level acoustic modeling with a stage-aware transformer for {ALS} dysarthria severity estimation: {ICASSP} 2026 {SAND} challenge,'' in \emph{ICASSP 2026--2026 IEEE International Conference on Acoustics, Speech and Signal Processing ({ICASSP})}.\hskip 1em plus 0.5em minus 0.4em\relax IEEE, 2026.

\bibitem{cho2024sylber}
C.~J. Cho, N.~Lee, A.~Gupta, D.~Agarwal, E.~Chen, A.~W. Black, and G.~K. Anumanchipalli, ``{Sylber}: Syllabic embedding representation of speech from raw audio,'' \emph{arXiv preprint arXiv:2410.07168}, 2024.

\bibitem{eyben2015geneva}
F.~Eyben, K.~R. Scherer, B.~W. Schuller, J.~Sundberg, E.~Andr{\'e}, C.~Busso, L.~Y. Devillers, J.~Epps, P.~Laukka, S.~S. Narayanan \emph{et~al.}, ``The {Geneva} minimalistic acoustic parameter set ({GeMAPS}) for voice research and affective computing,'' \emph{IEEE Transactions on Affective Computing}, vol.~7, no.~2, pp. 190--202, 2015.

\bibitem{eyben2013recent}
F.~Eyben, F.~Weninger, F.~Gross, and B.~Schuller, ``Recent developments in open{SMILE}, the munich open-source multimedia feature extractor,'' in \emph{Proceedings of the 21st ACM International Conference on Multimedia}, 2013, pp. 835--838.

\end{thebibliography}


\vfill\pagebreak

\end{document}